\documentstyle[prl,aps,twocolumn]{revtex}
\newcommand{\tr}{{\rm tr}}
\newcommand{\Tr}{{\rm Tr}}
\newcommand{\Dirac}{{\rm D}}

\newcommand{\eq}[1]{eq.~(\ref{eq:#1})}
\newcommand{\diag}{{\rm diag}}

\newcommand{\D}{D}

\newcommand{\bX}{\bar{X}}
\newcommand{\bM}{\bar{M}}
\newcommand{\bx}{\bar{x}}

\begin{document}

\draft
\tighten
\def\footnoterule{\kern-3pt \hrule width\hsize \kern3pt}

\title{Generalized heat kernel coefficients}

\author{L.L. Salcedo}

\address{
{~} \\
Departamento de F\'{\i}sica Moderna \\
Universidad de Granada \\
E-18071 Granada, Spain
}

\date{\today}
\maketitle

\thispagestyle{empty}

\begin{abstract}
Following Osipov and Hiller, a generalized heat kernel expansion is
considered for the effective action of bosonic operators. In this
generalization, the standard heat kernel expansion, which counts
inverse powers of a c-number mass parameter, is extended by allowing
the mass to be a matrix in flavor space. We show that the generalized
heat kernel coefficients can be related to the standard ones in a
simple way. This holds with or without trace and integration over
spacetime, to all orders and for general flavor spaces. Gauge
invariance is manifest.
\end{abstract}

\vspace*{0.2cm}

%\vspace*{0.2cm}

\vspace*{0.4cm}
%\vspace*{\fill}

%\newpage

Let the bosonic operator be
\begin{equation}
\Delta= -D^2+U(x),\qquad D_\mu= \partial_\mu+A_\mu(x)\,,
\end{equation}
where the gauge field $A_\mu(x)$ and the scalar field $U(x)$ are
matrices in some flavor space. This kind of operators appear
frequently in the computation of the effective action of fermions (see
e.g. \cite{Ball:1989xg}). In particular the normal parity component of
such an effective action is directly related to the determinant of the
operator $\Delta= \Dirac^\dagger\Dirac$ ($\Dirac$ being the Dirac
operator). A standard technical device to compute the effective action
is to split the scalar field into two contributions
\begin{equation}
U(x)= m^2+Y(x)
\end{equation}
where $m^2$ is a constant c-number (squared) mass parameter. This
allows to carry out an expansion in inverse powers of $m^2$ with
coefficients which are homogeneous polynomials constructed with the
quantities $Y$ and $D_\mu$. These coefficients are ordered by its
scale dimension and thus they are identical to those of the standard
heat kernel expansion
\cite{Schwinger:1951nm,Dewitt:1967ub,Seeley:1967ea}. These
coefficients are very well-known and we refer to
\cite{Ball:1989xg,vandeVen:1998pf} for details.

In the context of effective theories of quarks aiming at modeling QCD
at low energy \cite{Espriu:1990ff}, Osipov and Hiller
\cite{Osipov:2001th,Osipov:2001bj} consider instead a more general
separation of the scalar field
\begin{equation}
U(x)= M+Y(x)
\end{equation}
where $M$ is still a constant (i.e. $x$-independent) but not
necessarily a c-number. In general $M$ is a matrix in flavor space. In
a typical application $M=\diag(m_1^2,\ldots,m_n^2)$ for $n$ flavors,
$m_i$ being the constituent quark mass of the $i$-th flavor, and
$Y(x)$ accounts for the deviations of the scalar field from $M$.
Obviously in the particular case of degenerated masses the previous
case $M=m^2$ (a c-number) is recovered.

One can try to carry out an expansion in inverse powers of $M$ (i.e. a
large mass expansion) in this more general setting. A straight
approach is to organize the expansion in such a way that each term is
again a homogeneous polynomial in $D_\mu$ and $Y$. Technically this can
be done by introducing a bookkeeping parameter $\lambda$
\begin{equation}
\Delta_\lambda= M-\lambda D^2+\lambda Y(x) \,,
\end{equation}
and then proceed to expand the effective action in powers of
$\lambda$. This simple minded approach, however, meets the problem
that gauge invariance is not preserved separately by each term of the
expansion \cite{Osipov:2001th,Osipov:2001bj}. (Of course gauge
invariance holds for the full effective action functional.) The problem
is that we want to regard $\Delta$, as well as its associated
effective action, as a functionals where the external fields $A_\mu$
and $Y$ are the true variables and $M$ and the operator $\partial_\mu$
play the role of fixed parameters, i.e.
\begin{equation}
\Delta= \Delta(A,Y;M) = M- D^2+Y(x) \,.
\end{equation}
From this point of view, under a gauge transformation ($\Omega(x)$
being a matrix in flavor space)
\begin{equation}
\Omega^{-1}(x)\Delta(Y,A;M)\Omega(x)= \Delta(Y^\Omega,A^\Omega;M) \,.
\end{equation}
The quantities $M$ and
$\partial_\mu$ are gauge invariant by definition, $D_\mu$ and $U$
transform homogeneously under a similarity transformation and the external
fields $A_\mu$ and $Y$ transform inhomogeneously:
\begin{eqnarray}
\partial_\mu^\Omega &&= \partial_\mu\,,\quad
M^\Omega = M
 \nonumber \\ 
D_\mu^\Omega &&= \Omega^{-1}D_\mu\Omega\,,\qquad
U^\Omega = \Omega^{-1}U\Omega \\
A_\mu^\Omega &&= A_\mu+\Omega^{-1}[D_\mu,\Omega]\,,\qquad  
Y^\Omega= Y+\Omega^{-1}[U,\Omega]\,.  \nonumber
\end{eqnarray}
It is clear now that, because $\lambda$ does not affect $M$ in
$\Delta_\lambda$, the expansion in $\lambda$ breaks gauge invariance,
i.e. in general $\Delta_\lambda(Y^\Omega,A^\Omega;M)$ will not coincide with
$\Omega^{-1}\Delta_\lambda(Y,A;M)\Omega$ at $\lambda\not=1$. They do
coincide when $M=m^2$ is a c-number and in this case the standard
heat kernel expansion is recovered.

As an aside, we note that manifest gauge invariance is automatic order
by order in the context of a strict derivative expansion of the
effective action functional (e.g. \cite{Chan:1986jq}), that is
considering instead $\Delta_\lambda= U(x) - \lambda D^2$. The strict
derivative expansion can be viewed as a resummation of the heat kernel
expansion to all orders in $Y$. Recently explicit closed formulas have
been obtained in such an expansion for both the normal and the
abnormal parity components of the effective action of fermions coupled
to vector, axial, scalar and pseudo-scalar external fields and for an
arbitrary flavor group \cite{Salcedo:2001hp}. In the normal parity
case the formulas hold for arbitrary space-time dimension through
fourth order in the covariant derivatives. In the abnormal parity case
the leading order is computed in two and four dimensions.

The problem of obtaining a manifestly gauge invariant inverse mass
expansion for matricial $M$ has been solved by Osipov and Hiller in
\cite{Osipov:2001th,Osipov:2001bj} and explicit results are presented
there for lowest orders in the case of two and three flavors without
gauge fields. Presently we reformulate their approach in a way that
makes it simple to treat the case of arbitrary flavor group and the
introduction of gauge fields. Finally we find a simple relation
between the generalized heat kernel coefficients and the standard ones
so that no new calculation of these coefficients from scratch is
required.

In order to present the formalism let $X$ and $Z$ denote two matrices
(or operators) in some space $V$, such that the combination $X+Z$
transforms by a similarity transformation, $X$ represents the term
which is defined to be invariant under gauge transformations and $Z$
transforms inhomogeneously (thus, $X$, $Z$ and $X+Z$ generalize $M$,
$Y$ and $U$ respectively). Consider now the {\em gauge covariant}
quantity $f(X+Z)$ where $f(x)$ is some arbitrary function such as
e.g. the logarithm. (Note that $f$ itself is a c-number although its
argument, and thus its value, can be a matrix.) Loosely speaking what
we are seeking is to obtain an expansion for small $Z$ (or large $X$)
that generalizes the usual Taylor expansion valid for c-number $X$ and
$Z$ (or more generally, valid when $X$ and $Z$ commute), but in such a
way that each term of the expansion is separately gauge
covariant. This can be achieved as follows. Let the coefficients $Z_n$
be defined by the set of relations
\begin{equation}
(X+Z)^n = \sum_{k=0}^n \pmatrix{ n \cr k } \left\langle
X^{n-k} \right\rangle Z_k(Z,X)\,, \quad n= 0,1,2,\ldots
\label{eq:1}
\end{equation}
In this formula the notation $\langle A\rangle$ represents an average
of a matrix $A$ in $V$, namely
\begin{equation}
\langle A \rangle= \frac{\tr(A)}{\tr(1)} \,,
\end{equation}
where $\tr$ denotes the trace operation in $V$. The coefficients $Z_n$
are matrices and are recursively defined by the formula. The depend
both on $Z$ and $X$ in general. Of course when $X$ is a c-number $Z_n$
is simply $Z^n$.

Since $\{x^n,\ n=0,1,2,\ldots\}$ is a basis of functions we can take
linear combinations in the previous formula and write more generally
\begin{equation}
f(X+Z)= 
\sum_{n=0}^\infty
\frac{1}{n!} \left\langle f^{(n)}(X) \right\rangle Z_n \,,
\label{eq:1a}
\end{equation}
where $f(x)$ is an arbitrary function and $f^{(n)}(x)$ is its $n$-th
derivative.\footnote{As usual in quantum field theory, we will be
happy if the formulas hold in the sense of asymptotic series. They are
not required nor expected to be convergent.} Note that the
coefficients $Z_n$ do not depend on the function $f(x)$.

Three crucial properties of these coefficients can be established
without explicit computation:

(i) The $Z_n$ depend on $Z$ and $X$ but
this dependence is such that $Z_n$ remains unchanged if $X$ is
replaced by $X+a$ where $a$ is any c-number:
\begin{equation}
Z_n(Z,X)= Z_n(Z,X+a)\,.
\label{eq:11}
\end{equation}
This is because the shift introduced by $a$ can be absorbed by a
redefinition of $f(x)$.

(ii) The coefficients are gauge covariant since $(X+Z)^n$ is covariant
and $\langle X^n\rangle$ is invariant for all $n$:
\begin{equation}
Z_n(Z^\Omega,X)= \Omega^{-1}Z_n(Z,X)\Omega\,,\quad 
Z^\Omega= \Omega^{-1}(X+Z)\Omega-X\,.
\end{equation}

(iii) For the traced quantity
\begin{equation}
\tr\,f(X+Z)= 
\sum_{n=0}^\infty
\frac{1}{n!} \left\langle f^{(n)}(X) \right\rangle \,
z_n \,, \quad z_n=\tr(Z_n) \,.
\end{equation}
Then, the $z_n$ vanish for vanishing $Z$ (except $z_0$ which equals
$\tr(1)$):
\begin{equation}
z_n(0,X)= 0\,,\quad(n>0) \,.
\end{equation}

This latter property distinguishes this expansion from other possible
expansions which also enjoy the properties (i) and (ii), for instance
\begin{equation}
f(X+Z)= 
\sum_{n=0}^\infty
\frac{1}{n!} f^{(n)}(\langle X\rangle) \,
Z^\prime_n \,.
\end{equation}
The correct choice of $\langle f(X)\rangle$ among other possible
choices is a merit of \cite{Osipov:2001th,Osipov:2001bj}.

At lowest orders
\begin{eqnarray}
Z_0 &&= 1  \,, \nonumber \\ 
Z_1 &&=  Z+\bX \,, \\ 
Z_2 &&= (Z+\bX)^2 -\bx_2  \,, \nonumber \\ 
Z_3 &&= (Z +\bX)^3 -3\bx_2(Z+\bX) - \bx_3  \,, \nonumber 
\end{eqnarray}
where we have introduced the following notation
\begin{equation}
\bX= X-\langle X\rangle\,, \quad \bx_n= \langle \bX^n\rangle \,.
\end{equation}
For the traced coefficients
\begin{eqnarray}
z_0 &&= \tr(1)  \,, \nonumber \\ 
z_1 &&= \tr(Z) \,, \\ 
z_2 &&= \tr( Z^2 +2\bX Z ) \,, \nonumber \\ 
z_3 &&= \tr( Z^3 +3\bX Z^2 + 3 (\bX^2-\bx_2)Z) \,. \nonumber 
\end{eqnarray}
The property (i) noted above is manifest since $Z_n$ depends on $\bX$
only (and is independent of $\langle X\rangle $). Gauge covariance is
also obvious since $Z_n$ comes as a combination of powers of $Z+\bX$,
and this matrix transforms covariantly.

For subsequent application in the effective action problem some
properties of the $Z_n$ will be needed. First note that by taking a
first order variation with respect to $Z$, either in (\ref{eq:1}) or
(\ref{eq:1a}), and using the identity
\begin{equation}
\delta\,\tr(f(X+Z))= \tr(f^\prime(X+Z)\delta Z)\,,
\end{equation}
it follows that
\begin{equation}
\frac{\delta z_n}{\delta Z}= n Z_{n-1} \,,
\label{eq:2}
\end{equation}
which is well-known in the context of the heat kernel expansion
\cite{Ball:1989xg}. Next note that from their definition (\ref{eq:1})
and the property (\ref{eq:11}) above, which allows to use $\bX$
instead of $X$, one has to all orders
\begin{eqnarray}
Z_n &&= (Z+\bX)^n -\sum_{k=0}^{n-2} \pmatrix{ n \cr k } \bx_{n-k} Z_k 
 \label{eq:10a} \\
&&=  \sum_{k=0}^n\beta_{n,k}(Z+\bX)^k \label{eq:10}
\end{eqnarray}
for some c-number coefficients $\beta_{n,k}$ which do not depend on
$Z$. The allowable dependence of these coefficients on $n$ and $k$ can
be delimited by using (\ref{eq:2}) which immediately implies that
\begin{equation}
\beta_{n,k}=\frac{n}{k}\beta_{n-1,k-1}= \cdots=
\pmatrix{ n \cr k }\beta_{n-k,0} \,.
\end{equation}
Therefore (defining $\beta_n=\beta_{n,0}$) eq. (\ref{eq:10}) can be
given the following sharper form:
\begin{eqnarray}
Z_n=  \sum_{k=0}^n\pmatrix{ n \cr k }\beta_{n-k}(Z+\bX)^k \,.
\label{eq:3}
\end{eqnarray}
The recurrence (\ref{eq:10a}) then takes the form
\begin{eqnarray}
\beta_n &&= \delta_{n,0} -\sum_{k=2}^n
\pmatrix{ n \cr k } \bx_k \beta_{n-k}\,,
\label{eq:6}
\end{eqnarray}
and for lowest orders yields
\begin{eqnarray}
\beta_0 &&= 1  \,, \nonumber \\ 
\beta_1 &&= 0  \,,  \nonumber \\  
\beta_2 &&= -\bx_2  \,, \label{eq:7} \\
\beta_3 &&= -\bx_3  \,, \nonumber \\ 
\beta_4 &&= -\bx_4 +6\bx_2^2  \,, \nonumber
\end{eqnarray}
Note that this differs from the usual cumulant expansion beyond third
order.

A further identity will be needed to relate the standard and
generalized heat kernel coefficients. Let
\begin{equation}
\hat{T}_X= \sum_{r=0}^\infty \frac{\beta_r}{r!} 
\left(\frac{\partial}{\partial\langle X\rangle}\right)^r \,.
\label{eq:dT}
\end{equation}
Here $\partial/\partial \langle X\rangle$ refers to the dependence on
$\langle X\rangle$ of $f(X)$ written as $f(\bX+\langle X\rangle)$, and
so $\partial f(X)/\partial \langle X\rangle = f^\prime(X)$.  Then the
following identity holds
\begin{equation}
f(X+Z)= \hat{T}_X \sum_k \frac{1}{k!}\left\langle f^{(k)}(X)\right\rangle
(Z+\bX)^k\,.
\label{eq:T}
\end{equation}
This is easily proved as follows
\begin{eqnarray}
f(X+Z) &&= 
\sum_n
\frac{1}{n!} \left\langle f^{(n)}(X) \right\rangle Z_n
\nonumber \\
&&= 
\sum_{n,k}
\frac{1}{k!(n-k)!}\beta_{n-k} \left\langle f^{(n)}(X) \right\rangle (Z+\bX)^k
\nonumber \\
&&= 
\sum_{r,k}
\frac{1}{k!r!} \beta_r\left\langle f^{(k+r)}(X) \right\rangle (Z+\bX)^k
\nonumber \\
&&= \hat{T}_X \sum_k \frac{1}{k!}\left\langle f^{(k)}(X)\right\rangle
(Z+\bX)^k\,. 
\end{eqnarray}

Let us now turn to the application of the previous results to compute
the generalized heat kernel expansion. As is well-known the effective
action of the complex bosonic field with Klein-Gordon operator
$\Delta$ is $-\Tr(\log(\Delta))$ (where $\Tr$ refers to functional
trace). This and related functionals can be obtained once the ``current''
$\langle x| \Delta^{-1}|x\rangle$ is known.\footnote{It is understood
that the current is known for the whole family of operators $\Delta-\lambda$
for any complex $\lambda$, then
$$\Tr\,f(\Delta)= -\int d^dx\int_\Gamma \frac{d\lambda}{2\pi
i}f(\lambda) \tr\langle x| \frac{1}{\Delta-\lambda }|x\rangle \,$$
where the path $\Gamma$ encloses the spectrum of $\Delta$
\cite{Seeley:1967ea}.}  In the standard case when $M=m^2$ is a
c-number the current can be expanded as
\begin{equation}
\langle x| \Delta^{-1}|x\rangle
= \sum_{n=0}^\infty (-1)^n I_{n+1}a_n\,,
\label{eq:4}
\end{equation}
where
\begin{equation}
I_n= \int \frac{d^dp}{(2\pi)^d}\frac{1}{(m^2+p^2)^n}
\end{equation}
($d$ being the spacetime dimension) and $a_n$ are the (diagonal) heat
kernel coefficients. They are polynomials of dimension $2n$
constructed with $Y$, $F_{\mu\nu}=[D_\mu,\D_\nu]$ and their covariant
derivatives and they do not explicitly depend on $d$. At lowest orders
\begin{eqnarray}
a_0 &&= 1  \,, \nonumber \\
a_1 &&= Y  \,,   \\
a_2 &&= Y^2 -\frac{1}{3}[D_\mu,[D_\mu,Y]] 
+\frac{1}{6}F_{\mu\nu}F_{\mu\nu} \,. \nonumber
\end{eqnarray}
(Our convention is that of \cite{Avramidi:1991je,vandeVen:1998pf}
which differs from that of \cite{Ball:1989xg} by a factor $1/n!$)
The lowest order integrals $I_n$ are ultraviolet divergent and so some
renormalization is understood. Because the corresponding heat kernel
coefficients are polynomials, this renormalization translates into the
standard polynomial ambiguity in the effective action (and current
etc) in its ultraviolet divergent contributions.

Quite naturally, in the general case of arbitrary $M$ the generalized
heat kernel coefficients $b_n$ are defined as
\cite{Osipov:2001th,Osipov:2001bj}
\begin{equation}
\langle x| \Delta^{-1}|x\rangle
= \sum_{n=0}^\infty (-1)^n I_{n+1}b_n\,,
\label{eq:5}
\end{equation}
where now
\begin{equation}
I_n= \int\frac{d^dp}{(2\pi)^d}\left\langle \frac{1}{(M+p^2)^n}
\right\rangle
\end{equation}
and the average $\langle\ \rangle$ refers to flavor space.

Before embarking in the task of computing these generalized
coefficients from scratch, it is advisable to rest a moment and
consider what result is to be expected. The formulas regarding the
expansion of $f(X+Z)$ are fairly general (no assumption was made on
the vector space $V$), but of course they are formal due to
ultraviolet divergences when applied to the operator
$\Delta$. Technically a very definite problem in these calculations is
the lack of cyclic property of the trace when the operator $D_\mu$ (or
$\partial_\mu$) is involved. In order to avoid these complications,
let us temporarily neglect the contributions from derivatives. In this
case $\Delta$ equals $U=M+Y$ and this corresponds to $X=M$ and $Z=Y$
in the previous formulas. Expanding the current $(M+Y)^{-1}$ yields
$I_n= \langle (m^2)^{-n}\rangle$ and $a_n=Z^n=Y^n$ in the standard
expansion (\ref{eq:4}) and $I_n= \langle M^{-n}\rangle$ and $b_n=Z_n$
in the generalized case (\ref{eq:5}). $Z_n$ is given in \eq{3} as a
definite combination of $Z^n$ but with $Z$ shifted by $\bX$. In our
case $\bX$ corresponds to
\begin{equation}
\bM= M-\langle M\rangle \,.
\end{equation}
Therefore, for terms without derivatives, we obtain a simple relation
between standard and generalized heat kernel coefficients, which is
just a translation of \eq{3}, namely
\begin{eqnarray}
b_n=  \sum_{k=0}^n  \pmatrix{ n \cr k }\beta_{n-k}a^\prime_k
\label{eq:fun}
\end{eqnarray}
where $a^\prime_n$ denotes the usual heat kernel coefficient but using
everywhere
\begin{equation}
Y^\prime= Y+\bM
\end{equation}
instead of $Y$. In addition, the quantities $\beta_n$ are given by
the same formulas (\ref{eq:6},\ref{eq:7}) with $\bx_n= \langle
\bM^n\rangle$.

Because the simple relation (\ref{eq:fun}) is perfectly well-defined
and sensible also in presence of covariant derivatives it can be
conjectured that it holds in general. In fact this is the case, as
will be shown subsequently. This is our main result. To lowest orders
\begin{eqnarray}
b_0 &&= 1  \,, \nonumber \\
b_1 &&= Y+\bM  \,,   \\
b_2 &&= (Y+\bM)^2 -\frac{1}{3}[D_\mu,[D_\mu,Y+\bM]]
+\frac{1}{6}F_{\mu\nu}F_{\mu\nu} -\langle\bM^2\rangle\,. \nonumber
\end{eqnarray}

The analogous relation holds for the traced and integrated (over $x$)
coefficients needed for the effective action.  We have verified that
the results in \cite{Osipov:2001th} for the traced coefficients
$b_0,b_1,b_2,b_3,b_4$ in SU(2) are reproduced. We remark that the
replacement $Y\to Y^\prime=Y+\bM$ should be done everywhere in $a_n$
(i.e. in terms with derivatives too). The gauge covariance (in $Y$) is
obvious in $a^\prime_n$ since $Y^\prime$ is itself covariant. Another
remark is that the formula (\ref{eq:10}) is not sufficient to obtain
the result, in fact it does not even guarantee gauge invariance (in
$A_\mu$), and the more detailed formula (\ref{eq:3}) is needed.

Let us now turn to the proof of the relation (\ref{eq:fun}).  The main
observation is that we do not really need to compute the generalized
coefficients but only to relate them to the standard ones. Therefore
our strategy will be to start the computation of the coefficients and
at some point recognize that the relation (\ref{eq:fun}) will be
obtained.

There is an abundant literature on the computation of the heat kernel
coefficients in various settings \cite{vandeVen:1998pf}. Here we will
use a method convenient for our present purposes. The first step is to
use the method of symbols \cite{Salcedo:1996qy} to express the current
\begin{equation}
\langle x| \Delta^{-1}|x\rangle = \int \frac{d^dp}{(2\pi)^d} \langle
x|\frac{1}{(iD_\mu+p_\mu)^2+U} |0\rangle
\end{equation}
where $|0\rangle$ is the state with zero momentum $\langle
x|0\rangle=1$, and so $\partial_\mu|0\rangle=0$. This allows to deal
with the ultraviolet divergence $\langle x|x\rangle$ but explicit
gauge covariance is lost. Explicit gauge invariance is only recovered
after integration over $p_\mu$.

In order to apply our formulas, we identify
\begin{equation}
X=M+p^2\,,\quad Z=-D^2+2ip_\mu D_\mu+Y\,,
\end{equation}
regarded as operators in the space of position (spanned by
$|x\rangle$) and flavor. $p_\mu$ is a c-number parameter. Because $X$
is $x$-independent the averages $\langle f(X)\rangle$ or $\langle
f(\bX)\rangle$ are all in flavor space and well-defined. In particular
$\bX=\bM$. Further we define
\begin{equation}
J_n= \left\langle \frac{1}{(M+p^2)^n} \right\rangle\,,\quad
I_n= \int\frac{d^dp}{(2\pi)^d}J_n \,.
\end{equation}
A direct application of (\ref{eq:1a}) and (\ref{eq:T}) then gives
\begin{eqnarray}
\langle x|\Delta^{-1}|x\rangle &&= 
\int \frac{d^dp}{(2\pi)^d}
\sum_{n=0}^\infty (-1)^nJ_{n+1}\langle x|Z_n|0\rangle
\\
= \hat{T}_M &&
\int\frac{d^dp}{(2\pi)^d} \sum_{n=0}^\infty
(-1)^n J_{n+1} \langle x|(Z+\bM)^n|0\rangle \,.
\label{eq:12}
\end{eqnarray}
In this formula the operator $\hat{T}_M$ acts on the $J_n$. It is
given in (\ref{eq:dT}) and the $\beta_n$ are constructed with
$\bx_k=\langle\bM^k\rangle$.

A calculation of the coefficients would now proceed from (\ref{eq:12})
as follows (see e.g. \cite{Chan:1986jq}): (i) expanding the binomial,
(ii) using angular averaging in momentum space, (iii) using
integration by parts in momentum space (this step groups together
terms with a same dimension where it counts only the dimension carried
by $D_\mu$ and $Y^\prime=Y+\bM$ but not that carried by the momentum),
and (iv) bringing the expression to an explicit form where the
operators $D_\mu$ appear in covariant derivatives (i.e. in
commutators) only. These manipulations produce terms where all
operators are purely multiplicative (all derivative operators are
already inside commutators) and so equivalent to ordinary functions of
$x$, thus the matrix element $\langle x|\ |0\rangle$ simple evaluates
that function at $x$. This yields the heat kernel coefficients in this
approach. (We have explicitly computed $b_0$, $b_1$ and $b_2$ using
this method to verify that no subtleties arise.)  However this is not
necessary: it can be observed that when $M$ is a c-number $\bM$
vanishes and the formula becomes
\begin{eqnarray}
\langle x|\Delta^{-1}|x\rangle &&=
\int\frac{d^dp}{(2\pi)^d} \sum_{n=0}^\infty
(-1)^n J_{n+1} \langle x|Z^n|0\rangle \,.
\end{eqnarray}
All the manipulations (i-iv) just described can be carried out here
and we know that the final result is just the standard heat kernel
expansion quoted in (\ref{eq:4}). Then when these very manipulations
are used in (\ref{eq:12}) they will produce the same result except
that $Y$ is replaced by $Y+\bM$ (the fact that $J_n$ involves an
average over flavor does not make any difference). That is,
\begin{eqnarray}
\langle x|\Delta^{-1}|x\rangle &&=
\hat{T}_M \sum_{n=0}^\infty (-1)^n I_{n+1}a^\prime_n\,.
\end{eqnarray}
Finally, using
\begin{eqnarray}
\frac{\partial I_n}{\partial\langle M\rangle} = -n I_{n+1}
\end{eqnarray}
produces
\begin{eqnarray}
\langle x|\Delta^{-1}|x\rangle &&=
\sum_{k,n} (-1)^n I_{n+1} \pmatrix{ n \cr k }\beta_{n-k}a^\prime_k
\end{eqnarray}
and (\ref{eq:fun}) follows.

In summary, developing ideas put forward by Osipov and Hiller, we have
presented a general formalism to treat the problem of expanding
functionals above non c-number operators while preserving full gauge
invariance.\footnote{A similar construction has been considered in
the context of quantum gravity by Floreanini and Percacci
\cite{Floreanini:1992cw}.} We have shown that it is not restricted to
formal applications (finite dimensional spaces) since it holds too in
presence of ultraviolet divergences. This formalism has been applied
to obtain a simple relation (\ref{eq:fun}) between the standard and
the generalized heat kernel coefficients introduced in
\cite{Osipov:2001th,Osipov:2001bj}.

\section*{Acknowledgments}
I would like to thank C. Garc\'{\i}a Recio for comments on the
manuscript.  This work was supported in part by funds provided by the
Spanish DGICYT grant no. PB98-1367 and Junta de Andaluc\'{\i}a grant
no. FQM-225.

\end{document}